\documentstyle[12pt]{article}

\newlength{\dinwidth}
\newlength{\dinmargin}
\def\be{\nopagebreak[3]\begin{equation}}
\newcommand{\ee}{\end{equation}}
\def\ba{\begin{array}}
\def\ea{\end{array}}

\def\bea{\begin{eqnarray}}
\def\eea{\end{eqnarray}}

\begin{document}
\begin{titlepage}
\begin{flushright}
CERN-TH-7486/94\\
FTUAM-94/26\\
October, 1994\\
hep-th/9410237
\end{flushright}
\bigskip
\begin{center}
{\LARGE An Introduction to T-Duality \\in String
Theory}
\vskip 0.9truecm

{Enrique \'Alvarez
\footnote{Permanent address: Departamento de F\'{\i}sica
Te\'orica,
Universidad Aut\'onoma de Madrid, 28049 Madrid, Spain},
Luis \'Alvarez-Gaum\'e
and  Yolanda Lozano \footnotemark[1]}

\vspace{1pc}

{\em Theory Division CERN \\1211 Geneva 23\\ Switzerland}\\

\vspace{5pc}

{\large \bf Abstract}
\end{center}

In these lectures a general introduction to T-duality is given.
In the abelian case the approaches of Buscher, and Ro\u{c}ek and
Verlinde
are reviewed. Buscher's prescription for the dilaton transformation
is recovered from a careful definition of the gauge integration
measure. It is also shown how duality can be understood as a quite
simple canonical transformation.
Some aspects of non-abelian duality are also discussed,
in particular what is known on
relation to canonical transformations. Some implications of the
existence of duality on the cosmological
constant and the definition of distance in String Theory are also
suggested.

\vfill

\end{titlepage}

\tableofcontents
\newpage

\def\theequation{\thesection . \arabic{equation}}
\def\theequation{\thesubsection . \arabic{equation}}

\section{Introduction}
\setcounter{equation}{0}

Few words have been used with more different meanings than the
word ``duality''. Even within the restricted framework of string
theories, duality originally meant a symmetry between the s and
the t-channels in strong interactions (coming from the demands in
the S-matrix approach of the sixties of Regge behavior without fixed
poles and analiticity, which were shown to imply the existence of an
infinite number of resonances) \cite{scherk}. Somewhat related ideas,
also termed ``duality'', appear in the context of Conformal Field
Theory (CFT) as simple consequences of locality and associativity
of the operator product expansion (OPE) \cite{luger}.

Duality symmetry plays an important r\^ole in
Statistical Mechanics
(for a review and references to
the literature see for instance \cite{dizyk}), in particular in
the
analysis of the phase diagram of spin systems.  It can also be
understood as a way to show the equivalence between two
apparently different theories. On a
lattice system described by a Hamiltonian $H(g_i)$
with coupling constants $g_i$
the duality transformation
produces a new Hamiltonian $H^*(g^*_i)$ with coupling
constants $g^*_i$ on the dual lattice.  In
this way one can often relate the strong coupling regime
of $H(g)$ with the weak coupling regime of $H^*(g^*)$.
An important application was the determination
of the exact temperature at which the phase transition
of the two-dimensional Ising model takes place \cite{ising}.

More recently, the word ``duality'' (``space-time duality'') has
been introduced in yet another sense. T-duality is a
symmetry which relates physical
properties corresponding to big spacetime radius with quantities
corresponding to small radius. This will be our main theme
in this review and from now on we will refer to it as just duality
(a general reference is \cite{porrati}).
S-duality is a (conjectural) symmetry relating the strong coupling
regime with the weak coupling one, a bold generalization of the
original conjecture by Montonen and Olive \cite{montonen}.
Still more interesting
(and speculative), there is a ``duality of dualities'': S-duality
for strings corresponds to T-duality for fivebranes and
conversely (see \cite{sen}
for a general review). Another formally very similar property is
$\beta$-duality, a property of the free energy of strings at finite
temperature \cite{ot} which relates the
high and the low temperature phases.
For example, for the 10-dimensional heterotic string
\be
F(\beta)=\frac{\pi^2}{\beta^2}F(\pi^2/{\beta}).
\ee
The physical interpretation of this symmetry is, however, somewhat
uncertain due to the presence of the Hagedorn temperature.

In String Theory and Two-Dimensional Conformal Field Theory
duality is an important tool to show the equivalence of
different geometries and/or topologies and in determining
some of the genuinely stringy implications on the structure
of the low energy Quantum Field Theory limit.
Duality symmetry was first described on the
context of toroidal compactifications \cite{bgs}.
For the simplest case of a single
compactified dimension of radius $R$, the entire physics
of the interacting theory is left unchanged under the
replacement $R \rightarrow {\alpha}^{'} /R $
provided one also transforms the dilaton field $\phi
\rightarrow \phi - \log{(R/\sqrt{{\alpha}^{'}})}$
\cite{ema}. This simple case can
be generalized to  arbitrary toroidal compactifications
described by constant metric $g_{ij}$ and antisymmetric
tensor $b_{ij}$ \cite{nsw}. The
generalization of duality to this case becomes
$(g+b) \rightarrow (g+b)^{-1}$ and $\phi \rightarrow \phi
-{1\over 2}\log{\mbox{det}(g+b)}$. In fact this transformation
is an element of an infinite order discrete symmetry group
$O(d,d; Z)$ for $d$-dimensional toroidal compactifications
\cite{torodd,venezia1}. The
symmetry was later extended to the case
of non-flat conformal backgrounds
in \cite{buscher}.
In Buscher's construction one starts
with a  manifold $M$ with metric
$g_{ij}, i,j=0,\ldots d-1$, antisymmetric
tensor $b_{ij}$ and  dilaton field $\phi(x_i)$.
One requires the metric to admit at least one
continuous abelian isometry leaving invariant the
$\sigma$-model action constructed out of $(g, b, \phi)$.
Choosing an adapted coordinate system $(x^0, x^{\alpha}) =
(\theta, x^{\alpha}), \alpha = 1, \ldots d-1 $
where the isometry acts by translations of $\theta$, the
change of $g, b, \phi$ is given by
\bea
\label{busdual}
&&{\tilde g}_{00}=1/g_{00},\qquad
         {\tilde g}_{0\alpha}=b_{0\alpha}/g_{00},\nonumber\\
&&{\tilde g}_{\alpha\beta}=g_{\alpha\beta} -
(g_{0\alpha}g_{0\beta} - b_{0\alpha} b_{0\beta})/g_{00}\nonumber\\
&&{\tilde b}_{0\alpha} =g_{0\alpha}/g_{00},\nonumber\\
&&{\tilde b}_{\alpha\beta}=b_{\alpha\beta}-(g_{0\alpha}b_{0\beta}
         -g_{0\beta}b_{0\alpha})/g_{00},\nonumber\\
&&\tilde {\phi}=\phi-\frac12\log{g_{00}}.
\eea

The final outcome is that for any continuous isometry of the metric
which is a symmetry of the action one obtains the equivalence
of two apparently very different non-linear $\sigma$-models.
The transformation (\ref{busdual}) is referred to in the literature
as abelian duality due to the abelian character of
the isometry of the original $\sigma$-model.
If $n$ is the maximal number of commuting isometries, one gets
a duality group of the form $O(n,n;Z)$
\cite{givroc}.
Duality symmetries are useful in determining important
properties of the low-energy effective action, in particular in
questions related to supersymmetry breaking and to the lifting of
flat directions from the potential \cite{susydual}. Although the
transformation (\ref{busdual}) was originally obtained using
a method apparently
not compatible with
general covariance, it is not difficult to modify the
construction to eliminate this drawback \cite{aagbl}.
A particularly useful
interpretation of (\ref{busdual}) is in terms of the gauging
of the isometry symmetry
\cite{rocver}. The duality
transformation proceeds in two steps: i) First one
gauges the isometry group, thus introducing some auxiliary
gauge field variables $A$. The gauge field
is required to be flat and this is implemented by
adding a Lagrange multiplier term of the form
$\chi dA$.
It is naively clear that if we first perform the
integral ovel $\chi$, this provides a $\delta$-function
$dA$ on the measure, implying that
$A=dX$
is a pure gauge (we consider a
spherical world sheet for simplicity). Fixing $X=0$ the
original model is recovered. ii) The second step consists of
integrating first the gauge field $A$.
Since there is no gauge kinetic term, the integration
is gaussian, yielding a Lagrangian
depending on the
original variables and the auxiliary
variable $\chi$. After fixing the gauge the dual action follows.
In \cite{rocver} it was
further shown that if one starts with a conformal
field theory (CFT), conformal invariance is preserved
by abelian duality. The proof was based on an analogy
between the duality transformation
and the GKO construction \cite{r6}.

Of more recent history is the notion of non-abelian
duality \cite{quevedo,givnoab,r7,aagl}, which has no analogue in
Statistical Mechanics. The basic idea of \cite{quevedo},
inspired in the treatment of abelian duality
presented in \cite{rocver}, is to consider a conformal
field theory with a non-abelian symmetry group $G$.
In this case the gauge field
variables $A$ and the
Lagrange multipliers live in the
Lie algebra associated to $G$. The duality transformation
proceeds in the two steps described above.

In the abelian case it is also possible to work out
the mapping between some operators in the original and
dual theories, as well as the global topology of
the dual manifold \cite{aagbl}.
Thus for $G$
abelian we have a rather thorough understanding of
the detailed local and global properties of duality.
In the non-abelian case global information can only be
extracted for $\sigma$-models with chiral currents \cite{aagl}.
For these models it is possible to perform a non-local
change of variables in the Lagrange
multiplier term such that the Lagrangian keeps its local
expression and from it the global properties of the dual model
can be worked out. The same construction does not work for general
$\sigma$-models without chiral isometries.

Some interesting reviews on duality can be found in \cite{porrati}
and \cite{luis}.

\vspace*{0.5cm}

The organization of the lectures is as follows:
\begin{enumerate}

\item In section two we review the approaches of
Buscher \cite{buscher} and Ro\u{c}ek and Verlinde \cite{rocver}
to abelian duality. We also exhibit the kind of information one can
obtain with these formalisms. The approach of De la Ossa and
Quevedo \cite{quevedo} to non-abelian duality is explained.
Some comments are made
concerning the global properties
of the dual manifold \cite{aagl}.

\item In section three we show that for non-semisimple isometry
groups
a mixed gravitational-gauge anomaly may emerge in constructing
the non-abelian dual. This explains in particular why the
example considered in \cite{r7} violates conformal invariance
to first order in $\alpha^{'}$.

\item In section four we study with some detail the transformation
of the dilaton needed to preserve conformal invariance (to first
order in $\alpha^{'}$) under duality.

\item In section five the problem of the behavior of the
cosmological
constant under duality is addressed. This study is motivated
by the work
in \cite{horowitz} where an explicit example in which the
cosmological
constant changes under a duality transformation is considered.

\item In section six we study the implications that duality has
in the definition of a proper distance within String Theory.
We consider
particular families of correlators, manifestly duality invariant,
and discuss the properties a distance based on them would have.

\item In section seven the canonical transformation approach
to duality is studied. We shall be following \cite{aagl2}.
In the abelian case the explicit generating
functional producing Buscher's formulae is constructed.
It is shown that all the information which
can be obtained in the formulations above can
be derived more easily this way. The general formulation of
non-abelian
duality as a canonical transformation is so far unknown.
We review an example \cite{zachos} where a non-abelian
transformation in the $SU(2)$ principal chiral model
is constructed as a canonical transformation of type I, the same
type as for abelian duality.

\item Section eight contains a partial list of open problems.

\end{enumerate}

\section{Abelian and Non-Abelian Dualities}
\setcounter{equation}{0}

\subsection{Abelian Duality}
\setcounter{equation}{0}

We start with a summary of Buscher's formulation
\cite{buscher}.
Consider a non-linear $\sigma$-model defined on a
$d$-dimensional manifold $M$:
\be
\label{bus1}
S=\frac{1}{4\pi\alpha^{'}}\int d^2\xi
[\sqrt{h}h^{\mu\nu}g_{ij}\partial_\mu x^i\partial_\nu
x^j+i\epsilon^{\mu\nu}b_{ij}\partial_\mu x^i\partial_\nu
x^j+\alpha^{'}\sqrt{h}R^{(2)}\phi(x)],
\ee
where $g_{ij}$ is the target space metric, $b_{ij}$ the
torsion and $\phi$ the dilaton field, coupled to the two dimensional
scalar curvature in the world sheet $R^{(2)}$.
$h_{\mu\nu}$ is the world sheet
metric and
$\alpha^{'}$ the inverse of the string tension. Let us assume
that the
$\sigma$-model has an abelian isometry represented by a
translation in a coordinate $\theta$
in the target space. In the coordinates
$\{\theta, x^\alpha\},\,\alpha=1,\ldots , d-1$,
adapted to
the isometry, the metric, torsion and dilaton fields are
$\theta$-independent. Then the original theory can be obtained
from the following $d+1$-dimensional $\sigma$-model:
\bea
\label{bus2}
&&S_{d+1}=\frac{1}{4\pi\alpha^{'}}\int d^2\xi
[\sqrt{h}h^{\mu\nu}(g_{00}V_\mu V_\nu+2g_{0\alpha}V_\mu
\partial_\nu
x^{\alpha}+g_{\alpha\beta}\partial_\mu x^{\alpha}\partial_\nu
x^{\beta})
\nonumber\\
&&+i\epsilon^{\mu\nu}(2b_{0\alpha}V_\mu \partial_\nu
x^{\alpha}+b_{\alpha\beta}\partial_\mu x^{\alpha}\partial_\nu
x^{\beta})
+2i \epsilon^{\mu\nu}{\tilde \theta} \partial_\mu V_\nu
+\alpha^{'}\sqrt{h}R^{(2)}\phi(x)],
\eea
where $V$ is a 1-form defined on $M$ and
${\tilde \theta}$ is an additional variable acting as a Lagrange
multiplier. The equation of motion for
${\tilde \theta}$ implies
$\epsilon^{\mu\nu}\partial_\mu V_\nu=0$,
which in topologically trivial world sheets forces
$V_\mu=\partial_\mu \theta$, leading to the original theory.
If instead we integrate over the $V_\mu$-fields:
\be
\label{bus3}
V_\mu=-\frac{1}{g_{00}}(g_{0\alpha}\partial_\mu
x^{\alpha}+i\frac{\epsilon_\mu\,^{\nu}}{\sqrt{h}}(b_{0\alpha}
\partial_\nu
x^{\alpha}+\partial_\nu {\tilde \theta})),
\ee
we obtain the dual action:
\bea
\label{bus4}
{\tilde S}&=&\frac{1}{4\pi\alpha^{'}}\int d^2\xi
[\sqrt{h}h^{\mu\nu}({\tilde g}_{00}\partial_\mu {\tilde \theta}
\partial_\nu {\tilde \theta}
+2{\tilde g}_{0\alpha}\partial_\mu {\tilde \theta} \partial_\nu
x^{\alpha}+{\tilde g}_{\alpha\beta}\partial_\mu
x^{\alpha}\partial_\nu
x^{\beta}) \nonumber\\
&&+i\epsilon^{\mu\nu}(2{\tilde b}_{0\alpha}\partial_\mu
{\tilde \theta}\partial_\nu
x^{\alpha}+{\tilde b}_{\alpha\beta}
\partial_\mu x^{\alpha}\partial_\nu
x^{\beta})+\alpha^{'}\sqrt{h}R^{(2)}\phi(x)],
\eea
where:
\bea
\label{busch}
{\tilde g}_{00}&=&{1\over g_{00}} \nonumber\\
         {\tilde g}_{0\alpha}&=&{b_{0\alpha} \over g_{00}},
\qquad
{\tilde b}_{0\alpha}={g_{0\alpha} \over g_{00}} \nonumber\\
          {\tilde g}_{\alpha\beta} &=& g_{\alpha\beta} -
{g_{0\alpha}g_{0\beta} - b_{0\alpha} b_{0\beta}\over g_{00}}
\nonumber\\
        {\tilde
b}_{\alpha\beta}&=&b_{\alpha\beta}-{g_{0\alpha}b_{0\beta}
         -g_{0\beta}b_{0\alpha}\over g_{00}}.
\eea
(\ref{busch}) show that duality relates very different
geometries. We
will see that it may also lead to different topologies.
The integration on $V_\mu$ produces a factor in the measure
$\mbox{det}g_{00}$ which conveniently regularized yields
the shift
of the dilaton:
\be
\label{shifdil}
{\tilde \phi}=\phi-\frac12 \log{g_{00}}.
\ee
The regularization prescription in order to find (\ref{shifdil})
is fixed by requiring conformal invariance of the dual theory.
In \cite{buscher} the following definition was shown to yield
the correct dilaton shift satisfying conformal invariance to
first order in $\alpha^{'}$:
\be
\label{b1}
\mbox{det}A\equiv \frac{\mbox{det}\Delta_A}{\mbox{det}\Delta},
\ee
where
$\Delta_A=-\frac{1}{\sqrt{h}}\partial_{\mu}(\sqrt{h}h^{\mu\nu}A
\partial_{\nu})$,
$\Delta=-\frac{1}{\sqrt{h}}\partial_{\mu}(\sqrt{h}h^{\mu\nu}
\partial_{\nu})$. We will further justify this definition for the
determinant in section four.

The $\sigma$-model defined by
$({\tilde g}, {\tilde b}, {\tilde \phi})$ is independent of the
${\tilde \theta}$ variable, hence the original model can be
recovered
by performing the duality transformation with respect to
${\tilde \theta}$ shifts.

This formalism has apparently some limitations:

\begin{enumerate}

\item It seems that general covariance is broken due to the choice
of adapted coordinates needed to perform the duality
transformation.
This also obscures the issue of the global topology of the dual
manifold, which is harder to describe if one works in local
coordinates.

\item If the original theory has some isometries not commuting
with
the one used for duality they generically disappear as local
symmetries in the dual model.

\item The original model is recovered from (\ref{bus2}) only in
spherical world sheets. The monodromy of the $V$ variable
must be
fixed by imposing the absence of modular anomalies.
For that we need to know which are
the orbits of the Killing vector.

\item When the Killing vector has fixed points, $V_\mu$ in
(\ref{bus3}) is singular. In this case it could be much wiser
to work with the $d+1$-dimensional
action (\ref{bus2}).

\item What happens to the operator mapping from the above
construction?.

\item What are the general properties of the non-abelian
generalization?.

\end{enumerate}

All these questions can be addressed with a different way of
constructing the dual model. We will follow the work
of Ro\u{c}ek and Verlinde \cite{rocver}. The formulation of
Ro\u{c}ek and
Verlinde starts with the same $\sigma$-model (\ref{bus1}) with
the abelian isometry represented by
$\theta\rightarrow\theta +\epsilon$.
The key point is to gauge the isometry by introducing some
gauge fields $A_\mu$ transforming as
$\delta A_{\mu}=-\partial_{\mu}\epsilon$.
With a Lagrange
multiplier term the gauge field strength is required to
vanish,
forcing the constraint that the gauge field is
pure gauge. After gauge fixing the original model is then
recovered.

Gauging the isometry in (\ref{bus1}) and adding the Lagrange
multipliers term leads to:
\bea
\label{rv1}
S_{d+1}&=&\frac{1}{4\pi\alpha^{'}}\int d^2\xi
[\sqrt{h}h^{\mu\nu}(g_{00}(\partial_\mu\theta+A_\mu)
(\partial_\nu\theta+A_\nu)+2g_{0\alpha}(\partial_\mu
\theta +A_\mu) \partial_\nu
x^{\alpha} \nonumber\\
&&+g_{\alpha\beta}\partial_\mu
x^{\alpha}\partial_\nu x^{\beta})
+i\epsilon^{\mu\nu}(2b_{0\alpha}(\partial_\mu \theta +A_\mu)
\partial_\nu
x^{\alpha}+b_{\alpha\beta}\partial_\mu
x^{\alpha}\partial_\nu x^{\beta})
+2i \epsilon^{\mu\nu}{\tilde \theta} \partial_\mu A_\nu
\nonumber\\
&&+\alpha^{'}\sqrt{h}R^{(2)}\phi(x)].
\eea
The dual theory is obtained integrating the $A$ fields:
\be
\label{rv2}
A_\mu=-\frac{1}{g_{00}}(g_{0\alpha}\partial_{\mu}
x^{\alpha}+i\frac{\epsilon_\mu\,^{\nu}}{\sqrt{h}}(b_{0\alpha}
\partial_\nu x^{\alpha}+\partial_{\nu}{\tilde \theta})),
\ee
and fixing $\theta=0$.

In \cite{rocver} it is shown
that the original and dual theories can be considered as the
vectorial and axial cosets of a given higher dimensional
theory with
chiral currents in
which the abelian symmetry group is gauged. This shows that
conformal invariance is preserved by abelian duality to all
orders in $\alpha^{'}$ since one can think of the
initial and dual theories as two different functional integral
representations of the same conformal field theory.

Within this approach the open questions enumerated above can be
solved.

The procedure of gauging the isometry can be implemented in
arbitrary coordinates \cite{aagbl}. If the original
$\sigma$-model has a torsion term then Noether's procedure
must be followed, as made explicit in \cite{r11}.
Let us consider the following $\sigma$-model:
\bea
\label{eqnoriginal}
S&=& {1 \over  8\pi}\int g_{ij} \partial_{\mu} x^i
\partial^{\mu} x^j + {i \over 8\pi} \int b_{ij} dx^i
\wedge dx^j \nonumber\\
&=&{1 \over 2\pi} \int d^2\xi
(g_{ij} + b_{ij}) \partial x^i
{\overline \partial} x^j,
\eea
where $\alpha^{'}=2$.
Let $k^i$ be a Killing vector for the metric $g$:
\be
\label{killing}
{\cal L}_k g_{ij} = \nabla_i k_j + \nabla_j k_i =0.
\ee
Invariance of $S$ requires also
\be
\label{tururu}
{\cal L}_k b = d\omega, \,\,\,\,\,\,\,\, \omega =
i_k b - v,
\ee
where $(i_k b)_j\equiv k^i b_{ij}$ and $v$ is a one-form such that
$i_k H = -dv$ ($H=db$ locally).
The associated conservation law is:
\be
\label{lawcon}
\partial{\bar J}_k + {\bar \partial} J_k=0
\ee
\bea
\label{jotas}
J_k &=& (k - i_k b + \omega)_i \partial x^i =
(k - v)_i \partial x^i \equiv (k - v)\cdot \partial x \nonumber\\
{\bar J}_k &=& (k + i_k b - \omega)_i
{\bar \partial} x^i =
(k + v)_i {\bar \partial} x^i \equiv (k + v)\cdot
{\bar \partial} x.
\eea
If we wish to gauge the isometry we introduce gauge fields
$A, {\bar A}$, with $ \delta_{\epsilon}A = -\partial \epsilon
\,\, , \, \, \delta_{\epsilon}{\bar A} = -{\bar \partial}
\epsilon$, and $ \delta x^i = \epsilon k^i (x)$ now with
$\epsilon$ a function on the world sheet.
It can be shown \cite{aagbl} that the action:
\be
\label{dualac}
S_{d+1}={1\over 2\pi}\int d^2 \xi [(g_{ij}+b_{ij})\partial
x^i{\bar\partial}x^j+(J_k-\partial\chi){\bar A}
+({\bar J}_k+{\bar\partial}\chi) A + k^2 A
{\bar A}],
\ee
is invariant under:
\bea
\label{gaugetrans}
\delta_{\epsilon} x^i &=& \epsilon k^{i}(x) \qquad
\delta_{\epsilon} \chi = -\epsilon k \cdot v \nonumber\\
\delta_{\epsilon} A &=& -\partial \epsilon \qquad
\delta_{\epsilon} {\bar A} = -{\bar \partial} \epsilon .
\eea
The Lagrange multiplier term forces the gauge field to
be flat and at the same time cancels the anomalous variation
of the Lagrangian. For a genus $g$ world-sheet $\Sigma_g$
and compact isometry
orbits we may have large gauge transformations. We consider
multivalued gauge functions:
\be
\label{epsperiod}
\oint_{\gamma} d \epsilon = 2\pi n(\gamma) \qquad
n(\gamma) \in {Z},
\ee
where $\gamma$ is a non-trivial homology cycle in $\Sigma_g$.
Since we are dealing with abelian isometries it suffices to
consider only the toroidal case $g=1$. The variation of
$S_{d+1}$ is:
\bea
\label{Riemannid}
\delta S_{d+1} &=& {1\over 2\pi} \int \left( \partial
\chi {\bar \partial}\epsilon - \partial \epsilon {\bar
\partial} \chi\right) = {i\over 4\pi} \int_{T} d\chi \wedge
d\epsilon \nonumber\\
&=& {i\over 4\pi} \left( \oint_{a} d\chi \oint_{b} d\epsilon
- \oint_{a} d\epsilon \oint_{b} d\chi \right)
\eea
where $a$ and $b$ are the two generators of the homology
group of the torus T. Since $\epsilon$ is multivalued by
$2\pi {Z}$, we learn from (\ref{Riemannid}) that $\chi$ is
multivalued
by $4\pi {Z}$:
\be
\label{chiperiod}
\oint_{\gamma} d\chi = 4\pi m(\gamma)
\qquad m(\gamma) \in {Z}.
\ee
For a non-compact isometry $\delta S_{d+1} =0$ and $d\chi$ may
in general have real periods. The original theory is recovered
integrating the Lagrange multiplier, which appears in the action
in the form of a closed one form. In non trivial world sheets these
one forms have exact and harmonic components. The $\chi$-dependence
in (\ref{dualac}) is:
\be
\label{chi1}
S_\chi=-\frac{1}{2\pi}\int (d\chi_0+\chi_h)\wedge A.
\ee
Integrating by parts in the exact part and using Riemann's bilinear
identity we obtain:
\be
\label{chi2}
S_\chi=\frac{1}{2\pi}\int \chi_0\wedge dA
-\frac{1}{2\pi}(\oint_a \chi_h\oint_b A-
\oint_a A\oint_b \chi_h).
\ee
Integration on $\chi_0$ yields the constraint $dA=0$ and
integration on the harmonic components leads to:
\be
\oint_a A=\oint_b A=0.
\ee
Both constraints imply that $A$ must be an exact one form. Fixing
the gauge the original theory is recovered.
By construction $S_{d+1}$
is general covariant, and therefore we have a clear idea of
the $d$-dimensional
geometrical interpretation of the model. Locally
the dual manifold is equivalent to $(M/S^1)\times S^1$ (for compact
isometries), where the quotient means that the gauge is fixed
by dividing
by the orbits of the isometry group. Generically we expect topology
change as a consequence of duality. However
the more delicate issue is whether the dual manifold
${\tilde M}$ is indeed a product or a twisted product (non-trivial
bundle). It is  also useful to notice that in
the previous arguments
the structure of $\pi_{1}(M)$ played
no r\^ole. This rises some questions concerning the way the
operators in both theories are mapped under duality \cite{aagbl}.
The nature of the product relating the
gauged original manifold and the Lagrange multipliers space
turns out to be dictated by the gauge fixing procedure,
in particular
by Gribov problems. We use an example to labor
this point. This is the $SU(2)$ principal chiral model, which
represents a $\sigma$-model in $S^3$. The dual with respect to
a fixed point free abelian isometry is locally $S^2\times S^1$.
One knows
that this also holds globally when performing the gauge fixing.
This reveals that the dual manifold is $S^2\times S^1$ and not
a squeezed $S^3$ (for details see \cite{aagbl}).

The interest of working with the $d+1$-dimensional theory is that
the possible singularities in the dual theory due to the existence
of fixed points do not emerge. However if we are interested in
the explicit form of the dual $\sigma$-model we have to eliminate
the gauge field $A$. Integrating on $A$ in (\ref{dualac}),
the dual model $({\tilde g}, {\tilde b}, {\tilde \phi})$ reads:
\bea
\label{rokdual}
&&{\tilde g}_{00}={1\over k^2} \nonumber\\
        &&{\tilde g}_{0\alpha}={v_{\alpha} \over k^2}, \qquad
{\tilde b}_{0\alpha}={k_{\alpha} \over k^2} \nonumber\\
          &&{\tilde g}_{\alpha\beta}=g_{\alpha\beta} -
{k_{\alpha}k_{\beta} - v_{\alpha} v_{\beta}\over k^2}\nonumber\\
        &&{\tilde
b}_{\alpha\beta}=b_{\alpha\beta}-{k_{\alpha}v_{\beta}
        -k_{\beta}v_{\alpha}\over k^2} \nonumber\\
&&{\tilde \phi}=\phi-\frac12\log{k^2}.
\eea
Going to adapted coordinates and fixing the gauge
we recover Buscher's formulae since
$k^2=g_{00}$, $v_\alpha=b_{0\alpha}$, $k_\alpha=g_{0\alpha}$.
However this choice of coordinate system unables us to
obtain global information about the dual manifold.

The explicit operator mapping can be constructed \cite{aagbl}.
The duals to
vertex operators $V_p=\exp{ip\theta}$, which
are momenta operators in the direction
of the isometry, are non-local operators which
can be interpreted as winding operators
only for flat compact isometries. Thus, the description of
duality in
toroidal compatifications as the symmetry exchanging momenta and
windings is modified. In particular the structure of $\pi_{1}(M)$
turns out not to be important. The winding operators are associated
to compact isometry orbits in the target space manifold and not to
homologically non-trivial cycles as is
usually interpreted for toroidal compactifications.

The extension to non-abelian isometry groups is easily done in
this formalism. The details are worked out in the next section.

\subsection{Non-Abelian Duality}
\setcounter{equation}{0}

The same procedure \`a la Ro\u{c}ek and Verlinde was generalized
in \cite{quevedo} to
construct the dual with respect to a given no-abelian
isometry group $G$. The gauge fields take values in the Lie algebra
associated to the isometry group and they transform under gauge
transformations $x^m\rightarrow g^m\,_n x^n$, $m,n=1,\ldots,N$,
where $g\in G$, as $A\rightarrow g(A+\partial )g^{-1}$. The
isometry is gauged by introducing covariant
derivatives\footnote{Note that this way of gauging a continuous
global isometry is only valid for certain $\sigma$-models and
isometry
groups \cite{r11}.}:
\be
\label{nad1}
\partial x^m\rightarrow Dx^m=\partial x^m+A^\alpha
(T_\alpha)^m\,_n x^n,
\ee
where $T_\alpha$ is a $N$-dimensional representation for the
$\alpha$
generator of the Lie algebra of $G$.
The flatness of the gauge fields is imposed by the term:
\be
\label{nad2}
\int Tr (\chi F),
\ee
with $F=\partial {\bar A}-{\bar \partial}A+[A,{\bar A}]$.
The $\chi$-fields take values in
the Lie algebra associated to $G$ and transform in the adjoint
representation to preserve gauge invariance.
Integration on $\chi$ fixes $F=0$
in semisimple groups,
then $A$ is pure gauge (in spherical world sheets) and after gauge
fixing we recover the
original model.
As before the dual model is obtained integrating on $A$ and
then fixing the gauge.
For non-semisimple groups the Lagrange multipliers term must be
introduced in a different way since the Cartan-Killing metric is
degenerate and the integration on $\chi$ does not imply that all
the $F$-components are zero. In this case the $\chi$-fields must
be taken in the basis dual to $T_\alpha$ and they transform in the
coadjoint representation.

We can write the gauged $\sigma$-model action as:
\bea
\label{dloq5}
S_{\rm gauge}&=&\frac{1}{2\pi}\int
d^2z[Q_{mn}Dx^m {\bar D}x^n+Q_{m\mu}Dx^m {\bar \partial}x^{\mu}+
Q_{\mu n}\partial x^{\mu}{\bar D}x^n+Q_{\mu\nu}\partial
x^{\mu}{\bar \partial}x^{\nu} \nonumber\\
&&+Tr (\chi F)+\frac12 R^{(2)}\phi],
\eea
where $Q=g+b$, latin indices are associated to
coordinates adapted to the
non-abelian isometry and greek indices to inert
coordinates. We can write (\ref{dloq5}) as:
\be
\label{dloq7}
S_{\rm gauge}=S[x]+\frac{1}{2\pi}\int d^2z [{\bar
A}^{\alpha}f_{\alpha\beta}A^{\beta}+{\bar
h}_{\alpha}A^{\alpha}+h_{\alpha}{\bar
A}^{\alpha}+\frac12 R^{(2)}\phi],
\ee
with:
\bea
\label{dloq8}
h_{\alpha}&=&(Q_{mn}\partial x^m+
Q_{\mu n}\partial x^{\mu})
(T_{\alpha})^n\,_q
x^q-\partial
\chi^{\alpha}T_R\eta_{\alpha\alpha} \nonumber\\
{\bar h}_{\alpha}&=&(Q_{n\mu}{\bar \partial}
x^{\mu}+Q_{nm}{\bar \partial}x^m)(T_{\alpha})^n\,_q x^q+{\bar
\partial}\chi^{\alpha}T_R\eta_{\alpha\alpha} \nonumber\\
f_{\alpha\beta}&=&Q_{mn}(T_{\beta})^m\,_r (T_{\alpha})^n\,_p
x^r x^p+C_{\beta\alpha}\,^{\gamma}\chi^{\gamma}T_R
\eta_{\gamma\gamma},
\eea
where $[T_{\alpha},T_{\beta}]=C_{\alpha\beta}\,^{\gamma}
T_{\gamma}$
and $Tr (T_{\alpha}T_{\beta})=T_R \eta_{\alpha\beta}$
($Tr (T^{'}_{\alpha}T_{\beta})=T_R \eta_{\alpha\beta}$
if the group is not semisimple).

Integrating $A,{\bar A}$:
\be
\label{dloq9}
{\tilde S}=S[x]+\frac{1}{2\pi}\int d^2z [{\bar
h}_{\alpha}(f^{\alpha\beta})^{-1}h_{\beta}+\frac12 R^{(2)}
{\tilde \phi}],
\ee
where ${\tilde \phi}$ is given by:
\be
\label{dilanue}
{\tilde \phi}=\phi-\frac12\log{(\mbox{det}f)},
\ee
after regularizing the factor $\mbox{det}f$
coming from the measure as in previous section.
In all the examples considered the dual
model with this dilaton satifies
the conformal invariance conditions to first order in
$\alpha^{'}$,
but a general proof analogous to that of Buscher in the
abelian case is lacking.

The construction above seems to be a straightforward extension of
abelian duality. However this is not so. Non-abelian duality is
quite different from abelian duality, as it is clearly
manifested in
the context of Statistical Mechanics. In this context
duality transformations are applied to models defined
on a lattice $L$ with physical variables taking
values on some abelian group $G$.  The duality
transformation takes us from the triplet
$(L,G,S[g])$, where $S[g]$ is the action depending
on some coupling constants labelled collectively
by $g$ to a model $(L^*,G^*,S^*[g^*])$ on the dual
lattice $L^*$ with variables taking values on the
dual group $G^*$ and with some well-defined action
$S^*[g^*]$.  For abelian groups, $G^*$ is the
representation ring, itself a group, and when
we apply the duality transformation once again
we obtain the original model.  As soon as
the group is non-abelian the previous construction
breaks down because the representation ring
of $G$ is not a group \cite{dizyk}.
In particular the
non-abelian duality transformation
cannot be performed again to
obtain the model we started with.
In the context of String Theory the major problems in stating
non-abelian duality as an exact symmetry come when trying to
extend it to non-trivial world sheets and when performing
the operator mapping (a detailed explanation on this can be
found in \cite{aagbl}). With the usual Lagrange multipliers
variables is not possible to extract global information. In
$\sigma$-models with chiral currents
a non-local change of variables in the Lagrange multipliers
term can be done such that the dual Lagrangian is local.
In these variables the
dual theory can be shown to be the product of the coset of the
original manifold by the isometry group $M/G$
and the WZW model of group $G$ \cite{aagl}.
This applies in particular to the
case of abelian groups, in agreement with the results on
abelian duality. The dual variables introduced in \cite{aagl}
are the base of
the non-abelian bosonization
studied in \cite{quevedo2}.
As in the abelian case the more delicate issue is to know
what kind
of a product it is. This can be worked out in the case of
WZW models \cite{WZW}.
The partition function at genus one of a WZW model with group
$G$ is not a product of any modular invariant partition
function for the $G/H$ coset theory and one of the $H$
WZW-model (now $H$ is
the gauged isometry group). The Kac-Moody
characters of the $G_k$ WZW-model have a well-defined
decomposition in terms of products of $G/H$ and $H_k$
characters. This implies that the product is a twisted product.
In fact in this case the explicit integration on the $A$-fields
can be made and the result is that the model is self-dual.

For non-abelian isometry groups certain anomalies can arise
when performing the non-abelian dual construction
\cite{aagl,elitzur}.
When one analyzes carefully the measure of integration
over the gauge fields and its dependence on the world
sheet metric, one encounters a mixed gauge and
gravitational anomaly \cite{r8} when any generator
of the isometry group in the adjoint representation has
a non-vanishing trace.
This can only happen for non-semisimple groups.
This mixed anomaly generates a contribution to
the trace anomaly which cannot
be absorbed in a dilaton shift and imposes a mild anomaly
cancellation condition for the consistency of non-abelian
duality. We treat this point in the next section.

\section{Mixed Anomalies and Effective Actions}
\setcounter{equation}{0}

Since we
are interested in conformal invariance, we introduce an
arbitrary metric $h_{\alpha\beta}$ on the world sheet and
compute the contribution to the trace anomaly of the
auxiliary gauge fields $A_{\pm}^a$. If for simplicity
we work on genus zero surfaces, the most straightforward way to
compute the dependence of the effective action on the
world sheet
metric is to first parametrize $A_{\pm}$ as:
\be
\label{anom1}
A_+=L^{-1}\partial_+ L,\,\,\,A_-=R^{-1}\partial_-R,
\ee
for $L,R$ group elements. We can think of $x^{\pm}$ as
light-cone
variables or as complex coordinates, and they depend on the
metric being used. In changing variables from $A_{\pm}$ to
$(L,R)$ we encounter jacobians:
\be
\label{anom2}
{\cal D}A_+{\cal D}A_-={\cal D}L\,{\cal D}R\, \mbox{det}
(D_+(A_+)D_-(A_-))
\ee
with $A_{\pm}$ given by (\ref{anom1}) (we take $A_{\pm}$ as
antihermitian matrices). We can write the determinants in
(\ref{anom2}) in terms of a pair of (b,c)-systems
($b_{+a}, c^a$),
($b_{-a}, {\tilde c}^a$). $c,{\tilde c}$ are 0-forms
transforming in the adjoint representation of the group.
For arbitrary groups $b_{\pm}$ transform in the coadjoint
representation. The determinants in (\ref{anom2}) can be
exponentiated in
terms of the (b,c)- systems with an action:
\be
\label{anom3}
S[b_{\pm},c,{\tilde c}]=\frac{i}{\pi}\int (b_+D_-(A)c+
b_-D_+(A){\tilde c}),
\ee
which is formally conformal invariant. The variation of S with
respect to the metric is given by the energy-momentum tensor
$T_{\pm\pm}$. We can ignore momentarily that $A_{\pm}$ are
given by (\ref{anom1}) and work with arbitrary gauge fields.
We can compute the dependence of the effective action for
(\ref{anom3}) on the metric $h_{\alpha\beta}$ and the gauge
field using
Feynman graphs. Expanding about the flat metric, and using
the methods in \cite{r8}, the first diagrams contributing to
the effective action are

\let\picnaturalsize=Y
\def\picsize{1.0in}
\def\picfilename{diagrams.eps}
\ifx\nopictures Y\else{\ifx\epsfloaded Y\else\input epsf \fi
\let\epsfloaded=Y
\centerline{\ifx\picnaturalsize N\epsfxsize \picsize\fi
\epsfbox{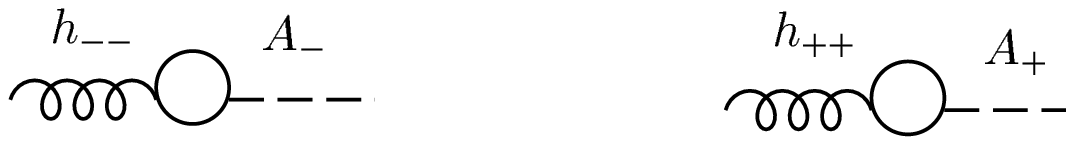}}}\fi

$h_{--} (h_{++})$ couples to $T_{++} (T_{--})$, and
$A_- (A_+)$ to the ghost currents $j_+ (j_-)$ given by:
\be
\label{anom4}
T_{++}=\partial_+c^a b_{+a}, \,\,\,
T_{--}=\partial_-{\tilde c}^a b_{-a}
\ee
\be
\label{anom5}
j_+^i=b_{+a}(T^i)^a_b c^b, \,\,\,
j_-^i=b_{-a}(T^i)^a_b {\tilde c}^b
\ee
If one keeps track of the $i\epsilon$ prescriptions in the
propagators appearing in the graphs, the loop
integrals are finite, and we can write their contributions
to the effective action as:
\be
\label{anom6}
W^{(2)}=\frac{1}{4\pi}Tr T^a \int d^2p(h_{--}(p)
\frac{p_+^2}{p_-}A_-^a(-p)
+h_{++}(p)\frac{p_-^2}{p_+}
A_+^a(-p)).
\ee
The coefficient of (\ref{anom6}) and $W^{(2)}$ may also be
computed
using the OPE:
\be
\label{ope}
T(z) j_a(w)\sim\frac{Tr T_a}{(z-w)^3}+
\frac{1}{(z-w)^2} j_a(w)
+\frac{1}{(z-w)}\partial j_a(w).
\ee
As it stands, $W^{(2)}$ has a gravitational anomaly, i.e. the
energy-momentum tensor is not conserved. However we can still
add local counterterms to (\ref{anom6}) to recover general
coordinate
invariance. Since to first order in $h$ the two-dimensional
scalar curvature has as Fourier transform:
\be
R(p)=2(2p_+p_-h_{+-}(p)-p_+^2h_{--}-p_-^2h_{++}),
\ee
if we add the counterterms:
\bea
&&W_{\rm c.t.}=\frac{1}{4\pi}Tr T_a \int A_-^a(-p)
(h_{++}(p)p_--
2p_+h_{+-})\nonumber\\
&&+\frac{1}{4\pi} Tr T_a \int A_+^a(-p)
(h_{--}(p)p_+-
2p_-h_{+-}),
\eea
we obtain an effective action
\be
\label{anom7}
W^{(2)}=\frac{1}{16\pi}Tr T_a  \int R(p)
\frac{p_+A_-^a(-p)+p_-A_+^a(-p)}{p_+p_-},
\ee
leading to a conserved energy-momentum tensor,
although it contains a trace anomaly which is not
proportional to $R(p)$ and therefore it cannot be absorbed in
a modification of the dilaton transformation.
Varying (\ref{anom7})
with respect to $h_{+-}$ leads to:
\be
\label{anom8}
\langle T_{+-}\rangle=\frac{\delta W^{(2)}}{\delta h_{+-}}=
\frac{1}{4\pi} Tr T_a (p_+A_-^a(-p)
+p_-A_+^a(-p))
\ee
which in covariant form becomes $\sim Tr T_a\, \nabla^{\alpha}
A_{\alpha}^a$.

Similarly we can vary the effective action to this order
with respect to gauge transformations to evaluate the
corresponding
gauge anomaly:
\bea
\label{anom9}
&&(D_- \frac{\delta W^{(2)}}{\delta A_-^a}+D_+ \frac{\delta
W^{(2)}}{\delta A_+^a})\sim p_-\frac{\delta W^{(2)}}
{\delta A_-^a(p)}+p_+\frac{\delta W^{(2)}}
{\delta A_+^a(p)} \nonumber\\
&&=p_- \langle j_{a+}(p)\rangle +
p_+\langle j_{a-}(p)\rangle
=-\frac{1}{8\pi}
Tr T_a R(-p).
\eea
This is a different way of writing the third order pole in the
OPE (\ref{ope}). From (\ref{anom8}) we see that at this
order ($W^{(2)}$) the trace
anomaly is not proportional to $R$, and it therefore cannot be
absorbed in a contribution to the dilaton or the effective
value of c (the central charge of the Virasoro algebra).
The contribution in (\ref{anom8}) spoils the conformal
invariance
of the dual theory, and further fields should be required
to cancel it. However in that case the resulting theory
would not agree with the one obtained through a naive
duality transformation. Another way to obtain the same
conclusion as in (\ref{anom8}) is
to use heat kernel methods. Both methods agree and we
conclude that the condition for the duality
transformation to respect conformal invariance is that the
generators of the duality group in the adjoint representation
should have a vanishing trace. The opposite may only happen
for non-semisimple groups, as in the example discussed
in \cite{r7}.

\section{The Transformation of the Dilaton}
\setcounter{equation}{0}

It is well known \cite{buscher} that the transformation
(\ref{busch})
is not the whole story. Indeed, the dual model
is not even conformally invariant in general, unless an
appropriate
transformation of the dilaton is included, namely
\be
\label{dil1}
{\tilde \phi}=\phi-\frac12\log{k^2}.
\ee
Perhaps the simplest way to realize that something has to
change in
the dilaton coupling
is to insist on the demand that the BRS charge be nilpotent.
It is well-known \cite{bns,peskin} that the BRS charge can be
written as:
\be
\label{dil2}
Q=\oint\frac{dz}{2\pi i}c(z)(T^{(x)}+\frac12 T_{gh}),
\ee
where
\bea
\label{dil3}
&&T^{(x)}\equiv -\frac12 g_{\mu\nu}\partial x^\mu \partial
x^\nu+\frac12 \partial^2\phi \nonumber\\
&&T_{gh}\equiv -2b\partial c-\partial b c
\eea
and $Q^2=0$ (in the OPE sense) is equivalent to the consistency
conditions of the $\sigma$-model $\beta$-functions
equal to zero \cite{callan}.

Using the fact that after performing a duality transformation
\be
\label{dil4}
{\tilde T}^{(x)}=T^{(x)}+\frac{1}{2k^2}[(k.\partial x)^2-
((v-w).\partial x)^2]+\frac12 \partial^2 ({\tilde \phi}-\phi)
\ee
the condition ${\tilde Q}^2=0$ necessarily leads to (\ref{dil1}).

We can trace the need for a transformation of the dilaton to the
behavior of the measure
under conformal transformations. Under a Weyl rescaling of the 2-d
world-sheet metric, $g\rightarrow e^\sigma g$, the integration
measure over the embeddings behaves (to first order in
$\sigma$) as:
\be
\label{dil5}
{\cal D}_{(e^\sigma g)}x={\cal D}_g x\, e^{\frac{d}{48\pi}
S_L(\sigma)+
6\alpha^{'}\int (-\nabla^2\phi+(\nabla\phi)^2-\frac14 R+
\frac{1}{48}H^2)\sigma},
\ee
where $S_L(\sigma)$ is the Liouville action. This means
that although
they are formally the same, both measures ${\cal D}x$ and
${\cal D}{\tilde x}$ behave in a very different way
under Weyl
transformations unless, of course, a compensating
transformation
of the dilaton is introduced to this purpose.

In the path integral approach the way to obtain the
correct dilaton
shift yielding to a conformally invariant dual theory can
be seen as
follows. Let us work with the approach of Ro\u{c}ek and
Verlinde. In complex coordinates and on spherical
world sheets we can parametrize $A=\partial\alpha$,
${\bar A}={\bar \partial}\beta$ (as we previously
did in (\ref{anom1})), for some 0-forms $\alpha, \beta$ in
the manifold $M$. The change of variables from $A, {\bar A}$
to $\alpha, \beta$ produces a factor in the measure:
\be
\label{reg1}
{\cal D}A {\cal D}{\bar A}={\cal D}\alpha {\cal D}\beta
(\mbox{det} \partial) (\mbox{det} {\bar \partial})=
{\cal D}\alpha {\cal D}\beta (\mbox{det}\Delta).
\ee
Substituting $A, {\bar A}$ as functions of $\alpha, \beta$
in (\ref{rv1}) and integrating on $\alpha, \beta$, the
following
determinant emerges:
\be
\label{reg2}
(\mbox{det}(\partial g_{00} {\bar \partial}))^{-1}.
\ee
In particular, the integration on $\beta$ produces a
delta-function
\be
\label{rec1}
\delta ({\bar \partial}(g_{00}\partial\alpha+
(g_{0\alpha}-b_{0\alpha})
\partial x^{\alpha}-\partial {\tilde \theta})),
\ee
which integrated on $\alpha$ yields the factor in the
measure
(\ref{reg2}).

What we finally get in the measure is then
\be
\label{reg3}
\frac{\mbox{det}\Delta}{\mbox{det}\Delta_{g_{00}}}
\ee
where $\Delta_{g_{00}}$ is given as in (\ref{b1})
in complex notation. This formula provides a justification
for
Buscher's prescription (see also \cite{st}) for the
computation
of the determinant arising from the naive gaussian
integration.
As we have just seen
some care is needed in order to correctly define the
measure of
integration over the gauge fields.
{}From (\ref{reg3}) the dilaton
shift (\ref{shifdil}) is obtained in the following way.
Writing $g_{00}$ as $g_{00}=1+\sigma\approx e^{\sigma}$ we have:
\be
\label{b3}
\Delta_{g_{00}}=(1+\sigma)\Delta-h^{\mu\nu}
\partial_{\mu}\sigma\partial_{\nu}.
\ee
Substituting in the infinitesimal variation of
Schwinger's formula:
\be
\label{b5}
\delta\log{\mbox{det}\Delta}=Tr \int_{\epsilon}^{\infty} dt
\delta\Delta
e^{-t\Delta}
\ee
we obtain
\be
\label{b6}
\delta\log{\mbox{det}\Delta_{g_{00}}}=-\int d^2\xi
\sqrt{h}\Omega
\langle\xi|e^{-\epsilon
(\Delta+\sigma\Delta-h^{\mu\nu}\partial_{\mu}
\sigma\partial_{\nu})}
|\xi\rangle,
\ee
where $\delta\Delta_{g_{00}}=-\Omega\Delta_{g_{00}}$
with $\delta
h_{\mu\nu}=\Omega \delta_{\mu\nu}$.
We can now use the heat kernel expansion \cite{gilkey}:
\be
\label{b7}
\langle\xi|e^{-\epsilon
D}|\xi\rangle=\frac{1}{4\pi\epsilon}+\frac{1}{4\pi}
(\frac16 R^{(2)}-V),
\ee
where
\be
\label{b8}
D\equiv
\Delta-2ih^{\mu\mu}A_{\mu}\partial_{\nu}+(-\frac{i}{\sqrt{h}}
\partial_{\mu}(\sqrt{h}h^{\mu\nu}A_{\nu})+
h^{\mu\nu}A_{\mu}A_{\nu})+V.
\ee
For
$D=\Delta+\sigma\Delta-h^{\mu\nu}\partial_{\mu}
\sigma\partial_{\nu}$
and after dropping the divergent term $1/4\pi\epsilon$
and the
quadratic terms in $\sigma$ we obtain:
\be
\label{b12}
\delta\log{\mbox{det}\Delta_{g_{00}}}=-\frac{1}{8\pi}
\int d^2\xi\sqrt{h}R^{(2)}\log{g_{00}}.
\ee
Substituting in (\ref{b1}):
\be
\label{final}
\mbox{det}g_{00}=\exp{(-\frac{1}{8\pi}\int
d^2\xi\sqrt{h}R^{(2)}\log{g_{00}})},
\ee
which implies ${\tilde \phi}=\phi-\frac12
\log{g_{00}}$.

\section{Duality and the Cosmological Constant}
\setcounter{equation}{0}

A striking feature of duality is the fact that the
cosmological constant, defined as the
asymptotic value of the scalar curvature, is not
in general invariant under the transformation. This fact
was first noticed in \cite{horowitz} for the case of a
WZW model with group $\widetilde{SL(2,R)}$ where a
discrete
subgroup was gauged.
This space has negative cosmological constant and under a
given
duality transformation it is mapped into an
asymptotically flat space (into a black string).
This implies that the usual definition
of the cosmological constant from the low-energy effective
action
is not satisfactory. Even at large
distances, if duality is not broken there is
a symmetry between local (momentum) modes
and non-local (winding) modes. One is lead to wonder to what
extent the cosmological constant
is a string observable\footnote{Similar remarks would apply
to the
concept of spacetime singularity in String Theory
\cite{witt,gins}.}.
The contribution
to the cosmological constant of the massless
sector might be cancelled by the tower of
massive states always present in String Theory
(proposals along these lines using the Atkin-Lehner
symmetry were advanced by G. Moore
\cite{moore}).

We study now the behavior of the scalar curvature
under duality.  If the space-time metric in the
$\sigma$-model takes the form
\be
\label{cosmoi}
ds^2=g_{ij}dx^idx^j\qquad
i,j =0,1,2,...,d-1,
\ee
where $x^0$ is adapted to the isometry
$\vec{k}=\partial /\partial x^0$, (\ref{cosmoi})
can be written as
\bea
\label{cosmoii}
ds^2&=&(e^0)^2+(g_{\alpha\beta}-{k_{\alpha}k_{\beta}
\over k^2})dx^{\alpha}dx^{\beta}\nonumber\\
e^0&=&kdx^0+{k_{\alpha}\over k}dx^{\alpha}\nonumber\\
k^2&=&k_ik^i=g_{00} \qquad k_{\alpha}=g_{0\alpha}.
\eea
Buscher's transformation leads to a dual
metric
\bea
\label{cosmoiii}
d{\tilde s}^2&=&({\tilde e}^0)^2+(g_{\alpha\beta}-
{k_{\alpha}k_{\beta}
\over k^2})dx^{\alpha}dx^{\beta}\nonumber\\
{\tilde e^0}&=&{1\over k}(d{\tilde x}^0
+v_{\alpha}dx^{\alpha}),
\eea
${\tilde x}^0$ being the Lagrange multiplier
and $v$ is defined as in section 2 by
$k^lH_{lij}=-\partial_{[i}
v_{j ]}, H=db$.  The dual scalar curvature
following from (\ref{cosmoiii}) is
\be
\label{cosmoiv}
{\tilde R}=R -{4\over k^2}g^{\alpha\beta}
\partial_{\alpha} k
\partial_{\beta} k +{4\over k}\Delta^{d-1}_q k+
{1\over k^2}H_{0\alpha\beta}H^{0\alpha\beta}-{k^2\over 4}
F_{\alpha\beta}F^{\alpha\beta},
\ee
where $\Delta^{d-1}_q$ is the $(d-1)$-dimensional
Laplacian for the metric
$g_{\alpha\beta}^q=g_{\alpha\beta}-
{k_{\alpha}k_{\beta}\over k^2}$, and
$F_{\alpha\beta}=\partial_{\alpha} A_{\beta}
 -\partial_{\beta} A_{\alpha}$
with $A_{\alpha}=k_{\alpha}/k^2$. (\ref{cosmoiv}) can be
rewritten as
\be
\label{cosmov}
{\tilde R}=R +4\Delta \log k +
{1\over k^2}H_{0\alpha\beta}H^{0\alpha\beta} -{k^2\over 4}
F_{\alpha\beta}F^{\alpha\beta}.
\ee
{}From (\ref{cosmov}) we see that:

\begin{itemize}

\item The only way to
``flatten'' negative curvature is by having
torsion in the initial space-time.  Otherwise
the dual of an asymptotically negatively curved space
time is a space of the same type.

\item Positive
curvature seems easier to flatten.

\item In general the
asymptotic behaviors of ${\tilde R}$ and $R$ are different,
which
proves the statement at the beginning of this section.

\item In the particular case of constant toroidal
compactifications ${\tilde R}=R$, in agreement with the
result in \cite{nair}.

\end{itemize}

We can also construct
the dual torsion
\bea
\label{cosmovi}
{\tilde H}_{0\alpha\beta}&=&-{1\over 2} F_{\alpha\beta}
\nonumber\\
{\tilde H}_{\alpha\beta\rho}&=&H_{\alpha\beta\rho}-
{3\over k^2}H_{0[\alpha\beta}k_{\rho]}-
{3\over 2}F_{[\alpha\beta}v_{\rho]}.
\eea
Since
\be
\label{cosmovii}
\sqrt{g}=k^2\sqrt{{\tilde g}},
\ee
and the modulus of $k$ can be expressed
in terms of the dilaton transformation properties,
\be
\label{cosmoviii}
{\tilde \phi}=\phi -\log k,
\ee
we obtain
\be
\label{cosmoix}
{\tilde R}+e^{2(\phi-{\tilde \phi})}
{\tilde H}^2_{0\alpha\beta} +\Delta {\tilde \phi}=
R+e^{2({\tilde \phi}-\phi)}H_{0\alpha\beta}^2+\Delta \phi,
\ee
which could be used to show the duality invariance
of the string effective action to leading order
in $\alpha '$.

The change of the cosmological constant under
duality is not only peculiar to three-dimensions
\cite{horowitz} but rather generic.  This raises the
physical question of whether in the context
of String Theory the value of the cosmological
constant can be inferred from the asymptotic
(long distance) behavior of the Ricci tensor.
If duality is not broken, the answer
seems to be in the negative, and
it makes the issue of what is the correct
meaning of the cosmological constant in
String Theory yet more misterious.

\section{The Physical Definition of Distance}
\setcounter{equation}{0}

The existence of duality raises the question of the
empirical definition of distance. This is, of course,
not a well defined
question in the absence of a sufficiently developed String
Field Theory, but can nevertheless be asked if we assume that
the outcome of every possible experiment is some correlation
function of the corresponding two dimensional CFT.

Thinking on the simplest situation of closed strings
propagating
in a spacetime with one coordinate compactified
in a circle, it is physically obvious that if we attempt to
measure distances through the asymptotic behavior of
correlation
functions at large separations\footnote{At large spatial
distances the propagator
behaves as $G(\vec{r}, \vec{r}^{'};M)\sim
e^{-M|\vec{r}-\vec{r}^{'}|}$
so that a suitable definition of distance is given by
$d(\vec{r},\vec{r}^{'})\equiv \frac{1}{M}
\log{\frac{G(\vec{r}, \vec{r}^{'};M)}{G(\vec{r},
\vec{r}^{'};2M)}}$.} \cite{david}, we would get a
completely
different answer if we use pure momentum states (of energy
$E_p=n/R$) or pure windings states (of energy $E_w=mR$),
which
we would most simply reinterpret as momentum states of
a torus
of radius $1/R$. This would lead us in a natural way to
restrict
the allowed outcome of our experiments to the
interval $d\in (1,\infty)$.

There are some technical complications, stemming from the
fact
that the Polyakov method only allows to compute on-shell
correlators, which means that we cannot probe directly
off-shell amplitudes.

In the absence
of any clear physical distinction among different classes of
states,
perhaps the most natural
possibility is to define distances out of ``unpolarized''
correlators, that is, considering the contribution of all states
at the same time.

There is still a certain freedom as to how to perform the
corresponding
Fourier transform in order to define physical quantities in
position
space. The most sensible thing seems, however, to make use of
the
fact that momentum and winding states define a
lattice \cite{narain,nsw}.
To be specific, sticking for concreteness to the case in which
$r$ dimensions (called $\vec{x}$) are compactified in circles
of
radius $R$, and
denoting by $\vec{y}$ the $(d-r)$-dimensional set of all
other
coordinates, the above considerations yield:
\be
\label{dist1}
G(\vec{x},\vec{x}^{'};\vec{y}-\vec{y}^{'},t-t^{'})
\equiv\sum_{\vec{n},\vec{m}}
e^{2\pi i(\vec{x}-\vec{x}^{'})(\vec{n}/R+\vec{m}R/2)}
\int dp_0 d\mu (\vec{p})
e^{i\vec{p}(\vec{y}-\vec{y}^{'})-ip_0(t-t^{'})}
\langle V_{\vec{n},\vec{m}} V_{-\vec{n},-\vec{m}}\rangle
f_{\vec{n},\vec{m}},
\ee
where $V_{\vec{n},\vec{m}}$ represents the vertex operator
corresponding to
the sector with momentum numbers $\vec{n}$ and winding numbers
$\vec{m}$,
and we will moreover consider pure solitonic states, without
any oscillators $N={\tilde N}=0$\footnote{In the unpolarized
case we are favouring, the selection function is trivial
$f_{\vec{n},\vec{m}}=1$, but we have written it in the formula
in order
to allow for more general possibilities.}.

The momentum space correlator is then given essentially by the
delta function implementing the condition that the vertex
operator
has conformal dimension 1, that is:
\be
\label{dist2}
p_0^2-\vec{p}^2-(\frac{\vec{n}}{R}+\vec{m}\frac{R}{2})^2=
N-1+{\tilde N}-1=-2.
\ee
A further restriction $(\vec{n}\vec{m}=0)$ comes from invariance
under
translations in $\sigma$ ($L_0={\tilde L}_0$).

Using the integral representation for the delta functions, the
integral over $p_0$ can be easily performed, and the double sum
packed into a Riemann theta-function:
\be
\label{dist3}
G(\vec{x},\vec{x}^{'};\vec{y}-\vec{y}^{'},t-t^{'})=\int
d\mu (\vec{p})
\int_{-\infty}^{\infty}d\tau \tau^{-1/2}
\int_{-\infty}^{\infty}
d\lambda e^{-i\frac{(t-t^{'})^2}{4\tau}-i\tau
(\vec{p}^2-2)+i\vec{p}
(\vec{y}-\vec{y}^{'})} \theta(\vec{z},\Omega),
\ee
where $\vec{z}\equiv (\vec{x}-\vec{x}^{'})
(\frac{R}{2},\frac{1}{R})$ and
\be
\label{dist4}
\Omega\equiv\frac{1}{\pi}\left(
\begin{array}{cc}
          -\tau R^2/2 & \frac{(\lambda-\tau)}{2} \\
          \frac{(\lambda-\tau)}{2} & -\frac{\tau}{R^2}
         \end{array}  \right).
\ee
A different expression can be obtained in terms of a
double sum
of $(d-r)$-dimensional Pauli-Jordan functions
\be
\label{dist5}
G^{(d)}(\vec{x},\vec{x}^{'};\vec{y}-\vec{y}^{'},t-t^{'})=
\sum_{\vec{m},\vec{n}}
G^{(d-r)}_{PJ}(\vec{y}-\vec{y}^{'},t-t^{'};
M^2(\vec{n},\vec{m}))
e^{2\pi i(\frac{\vec{n}}{R}+\vec{m}\frac{R}{2})
(\vec{x}-\vec{x}^{'})}\delta(\vec{n}\vec{m}),
\ee
where the ``mass spectrum'' is given by
\be
\label{dist6}
M^2(\vec{n},\vec{m})\equiv (\frac{\vec{n}}{R}+\vec{m}
\frac{R}{2})^2-2.
\ee

It is plain that any definition of distance based on the
preceding
ideas lacks any periodicity (which shows only in the
particular cases
in which pure winding states $f_{\vec{n},\vec{m}}=
\delta_{\vec{n},\vec{0}}$ or pure
momentum states $f_{\vec{n},\vec{m}}=
\delta_{\vec{m},\vec{0}}$ are used).

It is also arguable whether these correlators are indeed
the most natural ones to consider from the physical
point of view.
At extreme (either very high or very low) values of the
radius, ``pure''
states (winding or momentum) are much lighter than all
the others,
so that it is perhaps more natural to define distances
in terms of
the lightest states only \cite{branderva}.

One could always consider our suggestion as a
concrete implementation
of earlier speculations that at very short distances
there could
be a physical regime at which geometry ceases to be
smooth,
but distances can nevertheless be defined, and they obey
the triangular
inequality \cite{ecv}.

\section{The Canonical Approach}
\setcounter{equation}{0}

The procedures to implement duality explained in section 2
look unnecessarily complicated. In the
one due to Ro\u{c}ek and Verlinde the isometry is
gauged, the (non propagating) gauge fields are constrained
to be
trivial, and the Lagrange multipliers themselves are
promoted to
the rank of new coordinates once the gaussian integration
over
the gauge fields is performed. One suspects that all
those
complicated intermediate steps could be avoided, and
that it should
be possible to pass directly from the original to the
dual theory.

Some suggestions have indeed been made in the literature
pointing (at least in the simplified situation where
all backgrounds are constant or dependent only on time)
towards an understanding
of duality as particular instances of canonical
transformations \cite{venezia1,venezia2}.

In this section we are going to show that this idea
works well when the background admits
an abelian isometry \cite{aagl2},
laying duality on a simpler
setting than before, namely as a (privileged) subgroup of
the whole group of (non-anomalous, that is implementable
in Quantum Field Theory \cite{ghandour})
canonical transformations on the phase space of the theory.

We will
proof that Buscher's transformation formulae can be
derived by
performing a given canonical transformation on the Hamiltonian
of the
initial theory. We believe that this is a ``minimal''
approach in
the sense that no extraneous structure has to be introduced,
and all
standard results in the abelian case (and more) are easily
recovered
using it. In particular it is possible to perform the
duality
transformation in arbitrary coordinates not only in the
original
manifold (which was also possible in Ro\u{c}ek and
Verlinde's
formulation) but also in the dual one. The multivaluedness
and periods of the dual variables can be easily worked
out from the
implementation of the canonical transformation in the path
integral.
The generalization to arbitrary genus Riemann surfaces is in
this approach
straightforward.
The behavior of currents not commuting with those used
to implement duality can also be clarified. In the case
of WZW models it becomes rather simple to prove
that the full duality group is given by $Aut(G)_L\times
Aut(G)_R$, where $L,R$ refer to the left- and right-currents
on the model with group $G$, and $Aut(G)$ are the
automorphisms
of $G$, both inner and outer.  Due to the chiral
conservation
of the currents in this case, the canonical
transformation
leads to a local expression for the dual currents.
In the case where the currents are not chirally
conserved, then
those currents associated to symmetries not commuting
with
the one used to perform duality become generically
non-local
in the dual theory and this is why they are not
manifest
in the dual Lagrangian.
All the generators
of the full duality group $O(d,d;Z)$ can be described
in terms of canonical transformations.
This gives the impression that the duality group
should be understood in terms of global symplectic
diffemorphisms. It would be useful to formulate
it in the context of some analogue of the
group of disconnected diffeomorphisms, but for the
time being such a construction is lacking.

Concerning non-abelian duality, it seems to
fall beyond the scope of the Hamiltonian point
of view. There is one example \cite{zachos} in which
the non-abelian
dual has been constructed out of a canonical
transformation
but it is still early to say whether the general case can be
treated similarly.

\subsection{The Abelian Case}
\setcounter{equation}{0}

We start with a bosonic
sigma model written in arbitrary coordinates
on a manifold $M$ with Lagrangian
\be
\label{sigmamodel}
L = \frac12 (g_{ab}+b_{ab})(\phi) \partial_{+}\phi^a
\partial_{-}\phi^b
\ee
where $x^\pm=(\tau\pm\sigma)/2$, $a,b=1,\dots ,d={\rm dim}M$.
The corresponding Hamiltonian is
\be
H=\frac12 (g^{ab} (p_a-b_{ac}\phi^{'}\,^c)
(p_b-b_{bd}\phi^{'}\,^d) + g_{ab}\phi^{'}\,^a
\phi^{'}\,^b)
\ee
where $\phi^{'}\,^a\equiv d\phi^a/d\sigma$.
We assume moreover that there is a Killing vector
field $k^a$,
${\cal L}_k g_{ab}=0$ and $i_k H=-dv$ for some
one-form $v$,
where $(i_k H)_{ab}\equiv k^c H_{cab}$
and $H=db$ locally.
This guarantees the existence of a particular
system of coordinates, ``adapted coordinates'',
which we
denote by $x^i\equiv (\theta,x^\alpha)$, such that
$\vec{k}=\partial / \partial\theta$. We denote the
jacobian matrix by $e^i_a\equiv\partial x^i/
\partial \phi^a$.

This defines a point transformation
in the original Lagrangian (\ref{sigmamodel})
which acts on the Hamiltonian as a canonical
transformation
with generating
function $\Phi=x^i(\phi)p_i$, and yields:
\bea
\label{candual1}
&&p_a=e^i_a p_i \nonumber\\
&&x^i=x^i(\phi).
\eea
Once in adapted coordinates we can write the sigma model
Lagrangian as
\be
L=\frac12 G (\dot{\theta}^2-\theta^{'}\,^2)+(\dot{\theta}
+\theta^{'})J_-+
(\dot{\theta}-\theta^{'})J_++V
\ee
where
\bea
G=g_{00}=k^2 \qquad V=\frac12 (g_{\alpha\beta}+
b_{\alpha\beta})\partial_+
x^\alpha\partial_-x^\beta
\nonumber\\
J_-=\frac12 (g_{0\alpha}+b_{0\alpha})\partial_-x^\alpha
\qquad J_+=\frac12
(g_{0\alpha}-b_{0\alpha})\partial_+x^\alpha.
\eea
In finding the dual with a canonical
transformation we can use the Routh function with respect
to $\theta$,
i.e. we only apply the Legendre transformation to
$(\theta,\dot{\theta})$. The canonical momentum is given by
\be
p_\theta=G\dot{\theta}+(J_++J_-)
\ee
and the Hamiltonian
\bea
\label{hamil}
&&H=p_\theta \dot{\theta}-L=\frac12 G^{-1} p_\theta^2-
G^{-1}(J_++J_-)p_\theta+\frac12 G\theta^{'}\,^2+ \nonumber\\
&&+\frac12 G^{-1}(J_++J_-)^2+\theta^{'}(J_+-J_-)-V.
\eea
The Hamilton equations are:
\bea
\dot{\theta}=\frac{\delta H}{\delta p_\theta}=
G^{-1}(p_\theta-J_+-J_-) \nonumber\\
\dot{p_\theta}=-\frac{\delta H}{\delta\theta}=
(G\theta^{'}+J_+-J_-)^{'}
\eea
and the current components:
\bea
{\cal J_+}=\frac12 G \partial_+\theta+J_+=\frac12
p_\theta+\frac12
G\theta^{'}+\frac{J_+-J_-}{2} \nonumber\\
{\cal J_-}=\frac12 G \partial_-\theta+J_-=\frac12
p_\theta-\frac12
G\theta^{'}-\frac{J_+-J_-}{2}.
\eea
It can easily be seing that the current conservation
$\partial_-{\cal J_+}+\partial_+{\cal J_-}=0$ is equivalent
to the second Hamilton equation
$\dot{p_\theta}=-\delta H/ \delta\theta$.

The generator of the
canonical transformation we choose is:
\be
\label{fungen}
F=\frac12 \int_{D, \partial D=S^1} d\tilde{\theta}
\wedge d\theta=
\frac12 \oint_{S^1}
(\theta^{'}\tilde{\theta}-\theta\tilde{\theta}^{'})
d\sigma
\ee
that is,
\bea
\label{fungen2}
&&p_\theta=\frac{\delta F}{\delta\theta}=-
\tilde{\theta}^{'} \nonumber\\
&&p_{\tilde{\theta}}=-\frac{\delta F}{\delta\tilde{\theta}}
=-\theta^{'}.
\eea
This generating functional does not receive any quantum
corrections
(as explained in \cite{ghandour}) since it is linear
in $\theta$ and
$\tilde{\theta}$.  If $\theta$ was not an adapted
coordinate to
a continuous isometry, the canonical transformation would
generically lead to a non-local form of the dual Hamiltonian.
Since the Lagrangian and Hamiltonian in our case only depend
on the time- and space-derivatives of $\theta$, there are no
problems with non-locality. The transformation (\ref{fungen2})
in (\ref{hamil}) gives:
\bea
\label{canoni}
&&\tilde{H}=\frac12 G^{-1} \tilde{\theta}^{'}\,^2+
G^{-1}(J_++J_-)
\tilde{\theta}^{'}+ \nonumber\\
&&\frac12 G p_{\tilde{\theta}}^2-(J_+-J_-)
p_{\tilde{\theta}}+
\frac12 G^{-1} (J_++J_-)^2-V.
\eea
Since:
\be
\dot{\tilde{\theta}}=\frac{\delta\tilde{H}}{\delta
p_{\tilde{\theta}}}=Gp_{\tilde{\theta}}-(J_+-J_-),
\ee
we can perform the inverse Legendre transform:
\bea
&&\tilde{L}=\frac12 G^{-1} (\dot{\tilde{\theta}}^2-
\tilde{\theta}^{'}\,^2)+
G^{-1}J_+(\dot{\tilde{\theta}}-\tilde{\theta}^{'}) \nonumber\\
&&-G^{-1}J_-(\dot{\tilde{\theta}}+\tilde{\theta}^{'})+
V-2G^{-1}J_+J_-.
\eea
{}From this expression we can read the dual metric and torsion
and check that they are given by Buscher's
formulae\footnote{The
minus
signs in $\tilde{g}_{0\alpha}$ and
$\tilde{b}_{0\alpha}$
can be absorbed
in a redefinition $\tilde{\theta}\rightarrow -
\tilde{\theta}$.}:
\bea
\label{buscher}
&&\tilde{g}_{00}=1/g_{00},\qquad
         \tilde{g}_{0\alpha}=-b_{0\alpha}/g_{00},\qquad
          \tilde{g}_{\alpha\beta} = g_{\alpha\beta} -
\frac{g_{0\alpha}g_{0\beta} - b_{0\alpha} b_{0\beta}}{g_{00}}
\nonumber\\
&&\tilde{b}_{0\alpha} = -\frac{g_{0\alpha}}{g_{00}},\qquad
        \tilde{b}_{\alpha\beta}=b_{\alpha\beta}-
\frac{g_{0\alpha}b_{0\beta}-g_{0\beta}b_{0\alpha}}{g_{00}}
\eea
For the dual theory to be conformal invariant the dilaton
must transform as $\Phi^{'}=\Phi-\frac12\log{g_{00}}$
\cite{buscher} \cite{ema}. We have not been
able to find any
argument justifying this transformation within the canonical
transformations approach.

The dual manifold $\tilde{M}$ is automatically
expressed in coordinates adapted to the dual Killing vector
$\tilde{\vec{k}}=\partial / \partial\tilde{\theta}$.
We can now
perform
another point transformation, with the same jacobian as
(\ref{candual1})
to express the dual manifold in coordinates which are as
close as
possible to the original ones.

The transformations we perform are then: First a point
transformation $\phi^a\rightarrow \{\theta, x^\alpha\}$,
to go
to adapted coordinates in the original manifold. Then a
canonical transformation $\{\theta, x^\alpha\}\rightarrow
\{\tilde{\theta}, x^\alpha\}$,
which is the true duality transformation. And finally
another
point transformation $\{\tilde{\theta}, x^\alpha\}
\rightarrow
\tilde{\phi}^a$,
with the same jacobian as the first point transformation,
to express
the dual manifold in general coordinates.

It turns out that the composition of these three
transformations
can be expressed in geometrical terms using only the
Killing
vector
$k^a$, $\omega_a\equiv e^0_a$ and the corresponding dual
quantities\footnote{Note that we must raise and
lower indices
with the dual metric, i.e. $\tilde{e}_{ia}=\tilde{g}_{ij}
\tilde{e}^{j}_a, \tilde{e}^{ia}=\tilde{g}^{ab}
\tilde{e}^{i}_b$,
which implies $\tilde{\omega}_a=\omega_a$, but
$\tilde{\omega}^a=k^a (k^2+v^2)+\vec{e}\,^a
\cdot v$ (where
$\vec{e}\,^a\equiv e^a_{\alpha}$),
$\tilde{k}^a=k^a$ but
$\tilde{k}_a=(\omega_a -
(\vec{e}_a\cdot v))/k^2$. We have
moreover $\tilde{\omega}\,^2=
k^2+v^2+g^{\alpha\beta} v_{\beta}{\omega}_{\alpha}$ and
$\tilde{k}^2=1/k^2$.}.

It is then quite easy to check that the total canonical
transformation
to be made in (\ref{sigmamodel}) is just
\bea
\label{cancov}
&&k^a p_a \rightarrow \tilde{\omega}_a
\tilde{\phi}^{'}\,^a
\nonumber\\
&&\omega_a \phi^{'}\,^a \rightarrow \tilde{k}^a
\tilde{p}_a,
\eea
whose generating function is\footnote{The one-form
$\omega\equiv \omega_a d\phi^a$ is dual to
the Killing vector $\vec{k}$: $\omega(\vec{k})=1$,
$\omega(\vec{e}_\alpha)=0$, but it is of course
different from $\underline{k}\equiv k_a /k^2\, d\phi^a$
(the former is an exact form, whereas the latter does
not even in general satisfy Frobenius condition
$\underline{k}\wedge d\underline{k}=0$).}
\be
F=\frac12 \int_D \tilde{\omega}\wedge\omega=
\frac12 \int_D\tilde{\omega}_a d\tilde{\phi}^a\wedge
\omega_b d\phi^b.
\ee
One then easily performs the transformations in such a
way that
the dual metric and torsion can be expressed in
geometrical terms
as
\be
\tilde{g}_{ab}=g_{ab}-\frac{1}{k^2}(k_a k_b-
(v_a -\omega_a)(v_b-\omega_b))
\ee
\be
\tilde{g}^{ab}=g^{ab}+\frac{1}{(1+k.v)^2}
[(k^2+(v-\omega)^2)k^a k^b -2(1+k.v)(k^{(a}
(v-\omega)^{b)}]
\ee
and
\be
\tilde{b}_{ab}=b_{ab}-\frac{2}{k^2}k_{[a}
(v-\omega)_{b]},
\ee
where
\bea
k_{(a} (v-\omega)_{b)}=\frac12 (k_a(v_b-\omega_b)+
k_b(v_a-\omega_a))\nonumber\\
k_{[a} (v-\omega)_{b]}=\frac12 (k_a(v_b-\omega_b)-
k_b(v_a-\omega_a)).
\eea
These formulae are the covariant generalization of
(\ref{buscher}).
The canonical approach has been very useful in order
to obtain the dual manifold in an arbitrary coordinate system.
With the usual approaches it is expressed in
adapted coordinates to the dual isometry. This happens
because the
dual variables appear as Lagrange multipliers and after an
integration by parts only the derivatives of them emerge,
being then adapted coordinates automatically.

Some other useful information can be extracted easier in the
approach of the canonical transformation.

{}From the generating functional (\ref{fungen}) we can learn
about the multivaluedness and periods of the dual variables
\cite{aagbl}.
Since $\theta$ is periodic and in the path integral
the canonical transformation is implemented by \cite{ghandour}:
\be
\label{ghandour}
\psi_k[\tilde{\theta}(\sigma)]=N(k)\int
{\cal D}\theta(\sigma) e^{iF[\tilde{\theta},\theta(\sigma)]}
\phi_k[\theta(\sigma)]
\ee
where $N(k)$ is a normalization factor,
$\phi_k(\theta+a)=\phi_k(\theta)$ implies for $\tilde{\theta}$:
$\tilde{\theta}(\sigma+2\pi)-\tilde{\theta}(\sigma)=4\pi /a$,
which means that
$\tilde{\theta}$ must live in the dual lattice of $\theta$.
Note that (\ref{ghandour}) suffices to construct the
dual Hamiltonian.
It is a simple exercise to check that acting with
(\ref{canoni})
on the left-hand side of (\ref{ghandour}) and pushing
the dual
Hamiltonian through the integral we obtain the original
Hamiltonian acting on $\phi_k[\theta(\sigma)]$:
\be
\tilde{H}\psi_k[\tilde{\theta}(\sigma)]=N(k)\int {\cal D}
\theta(\sigma) e^{iF[\tilde{\theta},\theta(\sigma)]}
H\phi_k[\theta(\sigma)]
\ee
This makes the duality
transformation very simple conceptually,
and it also implies how it can be applied to arbitrary
genus Riemann surfaces, because the state
$\phi_k[\theta(\sigma)]$ could be the state obtained
by integrating the original theory on an arbitrary
Riemann surface with boundary.  It is also clear that
the arguments generalize straightforwardly when we
have several commuting isometries.

One can easily see that under the canonical transformation the
Hamilton equations are interchanged:
\bea
&&\dot{p_\theta}=-\frac{\delta H}{\delta\theta}=
(G\theta^{'}+J_+-J_-)^{'} \rightarrow
\dot{\tilde{\theta}}=Gp_{\tilde{\theta}}-J_++J_- \nonumber\\
&&\dot{\theta}=\frac{\delta H}{\delta p_\theta}=
G^{-1}(p_\theta-J_+-J_-) \rightarrow
\dot{p_{\tilde{\theta}}}=(G^{-1}(\tilde{\theta}^{'}+
J_++J_-))^{'},
\eea
and that
the canonical transformed currents conservation law is
in this case equivalent to the first Hamilton equation.

In the chiral case $J_-=0$ (i.e. $g_{0i}=-b_{0i}$) and
$G$ is
a constant, therefore we can
normalize $\theta$ to set
$G=1$ and :
\be
L=\frac12  (\dot{\theta}^2-\theta^{'}\,^2)+
(\dot{\theta}-\theta^{'})J_++V.
\ee
The Hamiltonian is
\be
H=\frac12 p_\theta^2-J_+p_\theta+\frac12 (J_++\theta^{'})^2-V.
\ee
The action is invariant under $\delta\theta=\alpha(x^+)$,
a $U(1)_L$ Kac-Moody symmetry.
The $U(1)$ Kac-Moody algebra has the
automorphism ${\cal J}_+\rightarrow -{\cal J}_+$. This is
precisely the effect of the canonical transformation.
The equation of motion or current conservation is:
\be
\partial_-(\partial_+\theta+J_+)=0.
\ee
${\cal J}_+=\partial_+\theta+J_+=p_\theta+\theta^{'}$ transforms
under the canonical transformation in
${\cal J}_+^{c.t.}=-\tilde{\theta}^{'}-
p_{\tilde{\theta}}=-{\cal J}_+$.

One can also follow the transformation to the dual model
of other continuous symmetries.  The simplest case is
as usual the WZW-model which is the basic
model with chiral currents.  Consider for simplicity
the level-$k$ $SU(2)$-WZW model with action
\be
S[g]={-k\over 2\pi}\int d^2 \sigma
Tr (g^{-1}\partial_+g g^{-1}\partial_-g)
+{k\over 12\pi}\int Tr(g^{-1}dg)^3.
\ee
The left- and right-chiral currents are
\be
{\cal J}_+={k\over 2\pi}\partial_+g g^{-1}\qquad
{\cal J}_-=-{k\over 2\pi}g^{-1}\partial_-g.
\ee
Parametrizing $g$ in terms of Euler angles
\be
g=e^{i\alpha\sigma_3/2}e^{i\beta\sigma_2/2}
e^{i\gamma\sigma_3/2},
\ee
${\cal J}_+$ are given by:
\bea
\label{corrientes}
&&{\cal J}^1_+={k\over 2\pi}(-\cos\alpha\sin\beta
\partial_+\gamma+\sin\alpha\partial_+\beta)\nonumber\\
&&{\cal J}^2_+={k\over 2\pi}(\sin\alpha\sin\beta
\partial_+\gamma+\cos\alpha\partial_+\beta)\nonumber\\
&&{\cal J}^3_+={k\over 2\pi}(\partial_+\alpha +
\cos\beta\partial_+\gamma),
\eea
and similarly for the right currents.
If we perform
duality with respect to $\alpha\rightarrow\alpha +
\mbox{constant}$,
${\cal J}^3_+\rightarrow- {\cal J}^3_+,
{\cal J}^3_-\rightarrow {\cal J}^3_-$ since ${\cal J}^3_+$
is the
current component adapted to the isometry.
For these currents it is
easy to find the action of the canonical transformation
because only the derivatives of $\alpha$ appear.  For
${\cal J}^{1,2}_+$ there is an explicit dependence on
$\alpha$
and it seems that the transform of these currents is very
non-local.  However due to its chiral nature,
one can show that there are similar chirally
conserved currents in the dual model. To do this we
first combine the currents in terms of root generators:
\bea
\label{root}
&&{\cal J}^{(+)}_+={\cal J}^1_+ +i{\cal J}^2_+=e^{-i\alpha}
(i\partial_+\beta-\sin\beta\partial_+\gamma)
=e^{-i\alpha}j^{(+)}_+ \nonumber\\
&&{\cal J}^{(-)}_+={\cal J}^1_+ -i{\cal J}^2_+=-e^{i\alpha}
(i\partial_+\beta+\sin\beta\partial_+\gamma)=
e^{i\alpha}j^{(-)}_+.
\eea
{}From chiral current conservation $\partial_-{\cal J}^{(\pm)}_+
=0$ we obtain
\be
\partial_- j^{(\pm)}_+=\pm i\partial_-\alpha j^{(\pm)}_+.
\ee
In these equations only $\dot\alpha,\alpha'$ appear, and
after the canonical transformation we can reconstruct the
dual non-abelian currents (in the previous equations
the canonical transformation amounts to the replacement
$\alpha\rightarrow\tilde{\alpha}$) which take the same
form as the original ones except that with respect to
the transformed ${\cal J}^3_+$ the r\^oles of positive and
negative roots get exchanged. One also verifies that
${\cal J}^a_-$ are unaffected.  This implies therefore
that the effect of duality with respect to shifts
of $\alpha$ is an automorphism of the current algebra
amounting to performing a Weyl transformation on the
left currents only while the right ones remain unmodified.
This result although known \cite{reviewkiritsis} is much
easier to derive in the Hamiltonian formalism
than in the Lagrangian formalism where one must
introduce external sources which carry some ambiguities.
The construction for $SU(2)$ can be straightforwardly
extended to other groups.  This implies that for
WZW-models the full duality group is
$Aut(G)_L\times Aut(G)_R$, where $Aut(G)$ is the group
of automorphisms of the group $G$, including Weyl
transformations and outer automorphisms.
For instance
if we take $SU(N)$, the transformation
$J_+\rightarrow -J_+^{T}$, i.e. charge conjugation,
follows from a canonical transformation of the
type discussed.  It suffices to take as generating
functions for the canonical transformation the sum
of the generating functions for each generator
in the Cartan subalgebra.
It is important
to remark that the chiral conservation of the
currents is crucial to guarantee the locality
of the dual non-abelian currents.  If the conserved
current with respect to which we dualize is not
chirally conserved locality is
not obtained.  The simplest example to verify
this is the principal chiral model for $SU(2)$,
which although is not a CFT serves for
illustrative purposes. The equations of motion for
this model imply the conservation laws:
\be
\partial_-{\cal J}^a_++\partial_+{\cal J}^a_-=0
\ee
where
\be
{\cal J}_\pm=\frac{k}{2\pi}\partial_\pm g g^{-1}.
\ee
If we perform duality with respect to the invariance under
$\alpha$ translations we know how ${\cal J}^3_\pm$
transform, since they are the currents associated to the
isometry. With the canonical transformation is possible
to see as well which are the other dual conserved currents.
Since the dual model is only $U(1)$-invariant
one expects the rest of the currents to become non-local
\cite{zachos}.
In terms of the root generators introduced in (\ref{root})
the conservation laws
\be
\partial_-{\cal J}^{(\pm)}_++\partial_+{\cal J}^{(\pm)}_-=0
\ee
are expressed:
\be
\partial_-j^{(\pm)}_++\partial_+j^{(\pm)}_-
\mp i(\partial_-\alpha j^{(\pm)}_++
\partial_+\alpha j^{(\pm)}_-)=0.
\ee
Performing the canonical transformation we obtain that
the dual conserved currents are given by:
\bea
&&{\tilde {\cal J}}^{(+)}_{\pm}=\exp{(i\int d\sigma
(\dot{\tilde{\alpha}}+\cos\beta \gamma^{'}))}
(i\partial_{\pm}\beta-\sin\beta\partial_{\pm}\gamma) \nonumber\\
&&{\tilde {\cal J}}^{(-)}_{\pm}=-\exp{(-i\int d\sigma
(\dot{\tilde{\alpha}}+\cos\beta \gamma^{'}))}
(i\partial_{\pm}\beta+\sin\beta\partial_{\pm}\gamma)
\eea
which cannot be expressed in a local form.

\subsection{The Non-Abelian Case}
\setcounter{equation}{0}

In view of the simplicity of the canonical approach to abelian
duality, one could be tempted to think that the corresponding
generalization to the non-abelian case would not be very
difficult. Unfortunately this is not the case, the reason
being that there are no adapted coordinates to
a set of non-commuting isometries, and therefore
one is led to a non-local form of the Hamiltonian.
In \cite{aagl} we could carry out the non-abelian
duality transformation due to the existence of
chiral currents and as a consequence of
the Polyakov-Wiegmann property \cite{polywieg}
satisfied
by WZW-actions.  Although in the intermediate steps
it was necessary to introduce non-local variables,
the final result led to a
local action in the new variables as a result
of the special properties of WZW-models mentioned.
The computations could be carried out exactly until
the end to evaluate the form of the effective
action in terms of the auxiliary variables needed
in the construction of non-abelian duals.
We have so far been unable to express
these functional integral manipulations in a Hamiltonian
setting as in the previous section.

To finish this section we present an example from the
literature in which a canonical transformation produces a
given non-abelian dual model. This example was
presented in \cite{zachos}. They consider the principal chiral model
with group $SU(2)$ and construct a local canonical transformation
mapping the model in a theory which turns out to be the
non-abelian dual with respect to the left action
of the whole group. This example was studied in the context
of non-abelian duality in \cite{fj,aagbl}.

The initial theory is the principal chiral model defined by the
Lagrangian:
\be
\label{nac1}
L=Tr (\partial_\mu g\partial^\mu g^{-1}),
\ee
where $g\in SU(2)$.
Parametrizing $g=\phi^0+i\sigma^j\phi^j$,
with $\phi^0,\phi^j$ subject
to the constraint $(\phi^0)^2+\phi^2=1$ and $\phi^2\equiv
\sum_j (\phi^j)^2$, (\ref{nac1}) becomes:
\be
\label{nac2}
L=\frac12 (\delta^{ij}+\frac{\phi^i\phi^j}{1-\phi^2})
\partial_\mu\phi^i\partial^\mu\phi^j.
\ee
The generating functional:
\be
\label{nac3}
F[\psi,\phi]=\int_{-\infty}^{+\infty} dx \psi^i
(\sqrt{1-\phi^2}\frac{\partial}{\partial x}\phi^i-\phi^i
\frac{\partial}{\partial x}\sqrt{1-\phi^2}+
\epsilon^{ijk}\phi^j\frac{\partial}{\partial x}\phi^k)
\ee
produces the canonical transformation:
\bea
\label{nac4}
p_i&=&\frac{\delta F[\psi,\phi]}{\delta\psi^i}=(\sqrt{1-\phi^2}
\delta^{ij}+\frac{\phi^i\phi^j}{\sqrt{1-\phi^2}}-
\epsilon^{ijk}\phi^k)\frac{\partial}{\partial x}\phi^j
\nonumber\\
{\tilde p}_i&=&-\frac{\delta F[\psi,\phi]}{\delta\phi^i}=
(\sqrt{1-\phi^2}\delta^{ij}+\frac{\phi^i\phi^j}
{\sqrt{1-\phi^2}}
+\epsilon^{ijk}\phi^k)\frac{\partial}{\partial x}\psi^j+
\nonumber\\
&&(\frac{2}{\sqrt{1-\phi^2}}(\phi^i\psi^j-\psi^i\phi^j)-
2\epsilon^{ijk}\psi^k)\frac{\partial}{\partial x}\phi^j,
\eea
which transforms (\ref{nac2}) into:
\be
\label{nac5}
{\tilde L}=\frac{1}{1+4\psi^2}[\frac12 (\delta^{ij}+4\psi^i
\psi^j)\partial_\mu\psi^i\partial^\mu\psi^j-\epsilon^{\mu\nu}
\epsilon^{ijk}\psi^i\partial_\mu\psi^j\partial_\nu\psi^k].
\ee
This is the non-abelian dual of (\ref{nac1}) with respect to
the left action of the whole group \cite{fj,aagbl}.
The generating functional (\ref{nac3}) can be written as:
\be
\label{nac6}
F[\psi,\phi]=\int_{-\infty}^{+\infty}dx \psi^i J_i^1[\phi]
\ee
where $J_i^1[\phi]$ are the spatial components of the
conserved
currents of the initial
theory. (\ref{nac6}) is linear in the dual variables but not in
the initial
ones, so it will receive quantum corrections when implemented
in the path integral. From (\ref{nac4}) it is not obvious
that the
dual model will not depend on the original variables $\phi^i$.
However this is so. Whether this way of constructing the
generating
functional of non-abelian duality is general or only works in
this
particular example is still an open question.

\section{Conclusions and Open Problems}
\setcounter{equation}{0}

In these lectures a general exposition of abelian and
non-abelian
duality has been given. The usual approaches in the literature
to both kinds of dualities
have been reviewed. The goals of these
approaches have been also exhibited, and some of
them derived explicitly, as the formulation of abelian duality
in an
arbitrary coordinate system. The canonical transformations
approach to
abelian duality presented in \cite{aagl2} has been studied in
detail,
focussing especially in the problems that could not be
solved in the
usual approaches of Buscher or Ro\u{c}ek and Verlinde or were
difficult to study. The non-abelian case has been also
considered,
although the general construction as a canonical
transformation is
not yet understood. As was mentioned
in the lectures the example
given by Curtright and Zachos in \cite{zachos} opens the
possibility for non-abelian duality to be formulated in this
way, in spite of the difficulties already mentioned concerning
the impossibility of finding an adapted coordinate system to
the whole set on non-commuting isometries.

The relation between duality and external automorphisms
\cite{bs,ao,ko} is also
much in need of further clarification.

\bigskip

{\large\bf Acknowledgements}

We would like to thank C. Zachos for useful suggestions.
One of us (LAG) would like to thank the organizers of
the Trieste Spring
School for the opportunity to present this material and
for their
kind hospitality. E.A. and Y.L. were supported in part by
the CICYT
grant AEN 93/673 (Spain) and by a fellowship from Comunidad
de
Madrid (YL). They would also like to thank the Theory
Division at
CERN for its hospitality while part of this work was
done.

\end{document}